\documentclass[reprint,a4paper,australian,amsmath,aps,prd,twocolumn,noshowpacs]{revtex4-1}

\usepackage{amsmath}
\usepackage{enumerate}
\usepackage{amssymb}
\usepackage{amsfonts}
\usepackage{graphicx}
\usepackage{bm}
\usepackage{subfigure}
\usepackage{amscd}
\usepackage{epsfig}
\usepackage{hhline,multirow}
\usepackage{dcolumn}
\usepackage{url}
\usepackage[dvipsnames]{xcolor}
\definecolor{DarkBlue}{rgb}{0.0, 0.0, 0.4}
\usepackage[colorlinks=true,linkcolor=blue,urlcolor=DarkBlue]{hyperref}
\usepackage{indentfirst}
\usepackage{psfrag}
\usepackage{slashed}
\usepackage{float}
\usepackage{setspace}
\usepackage{braket}

\begin{document}

\title{Sea-quark loop contributions to the $\bar d - \bar u$ asymmetry in the proton}

\author{Derek B. Leinweber}
\affiliation{Centre for the Subatomic Structure of Matter (CSSM), 
	     Department of Physics, University of Adelaide, SA, 5005, Australia}

\author{Anthony W. Thomas}
\affiliation{Centre for the Subatomic Structure of Matter (CSSM), 
	     Department of Physics, University of Adelaide, SA, 5005, Australia}

\begin{abstract}
QCD interactions for equal-mass fermion flavors are flavor blind. This fact is often used to
state that disconnected sea-quark loop contributions are equal for $u$ and $d$ quarks in the mass
symmetric case and therefore these disconnected sea-quark loop contributions cannot contribute to
the well-known $\bar d - \bar u$ asymmetry in the proton.  Instead, it is argued that one must look
to the connected sector of lattice QCD correlation functions to find this difference. In this
presentation, we note that these statements are true provided unphysical contributions in the
sea-quark loop sector are included, contributions from baryons that do not appear in the physical
spectrum.  To respect the Pauli principle, these unphysical contributions from the disconnected
sea-quark loop sector must cancel equally unphysical contributions in the connected sector.  The
remaining disconnected sea-quark loop contributions no longer have a balance between $\bar d$ and
$\bar u$.  Upon considering only physically observed baryons in the loop contributions, we
illustrate an important contribution from the sea-quark loop sector to $\bar d - \bar u$ that
enhances the leading connected contribution by 12\%.
\end{abstract}


\maketitle

\section{Introduction}
Understanding the structure of the nucleon remains an exciting challenge for modern nuclear and
particle physics \cite{Gross:2022hyw}. One of the great surprises of the past few decades in this
area was the discovery that there is an asymmetry between the anti-down ($\bar d$) and anti-up
($\bar u$) sea quarks in the proton. While this was in fact predicted on the basis of chiral
symmetry~\cite{Thomas:1983fh}, it was almost a decade later that a hint of such an
asymmetry~\cite{Ito:1980ev} was verified experimentally by the New Muon
Collaboration~\cite{NewMuonNMC:1990xyw}, as a significant discrepancy in the Gottfried
sum-rule~\cite{Gottfried:1967kk,Gottfried:1986hc}.

Since then, the shape of $\bar{d} - \bar{u}$ as a function of Bjorken-$x$, has been studied in
detail at leading twist through the Drell-Yan process~\cite{SeaQuest:2021zxb,NuSea:2001idv} as well
as further measurements of $F_{2}^p \, - \, F_{2}^n$~\cite{NewMuon:1993oys}. There have also been
extensive studies of the capacity of chiral field theory to constrain this and other asymmetries in
the sea of the nucleon~\cite{Thomas:2000ny,He:2022fne,Kretzer:2003wy,Salamu:2019dok,Wang:2016ndh,%
  Burkardt:2012hk,Alberg:2017ijg,Diakonov:1996sr,Dressler:1999zv,Melnitchouk:1991ui,Henley:1990kw,Signal:1987gz}.

Apart from the intrinsic interest in nucleon structure, modern tests of the Standard Model demand
that the parton distribution functions (PDFs) be known very
accurately~\cite{NNPDF:2021njg,AbdulKhalek:2022hcn,Thomas:2021lub,NNPDF:2017mvq,Zheng:2021hcf,Bentz:2009yy,Martin:2009iq}.
Such considerations have led to interesting suggestions that constraints from lattice QCD might
supplement experimental data in constraining hadronic
PDFs~\cite{Detmold:2020snb,JeffersonLabAngularMomentumJAM:2022aix,Hou:2022ajg,Liu:2012ch}.

In describing the structure of the proton in the context of deep inelastic scattering and parton
distribution functions, the convention is to distinguish between a three-quark valence sector and a
quark-number zero sea sector consisting of quark-antiquark pairs and gluons \cite{Zhang:2001gna}.
Here our focus is on the connected and disconnected quark flows considered in lattice QCD
calculations of three-point correlation functions, where a flavor-diagonal quark-bilinear current
probes the structure. While the disconnected sector generates quark-number zero sea contributions,
the connected sector contributes to both the three-quark valence- and the zero-quark sea-quark
contributions, due to the ability of the connected quark flows to include additional
quark-antiquark sea contributions.  The relationship between the valence and sea sectors of
phenomenology and the connected and disconnected sector of quantum field theory was established
long ago \cite{Liu:1993cv,Liu:1999ak}.

Here we focus on suggestions that the difference in the sea quark contributions from connected and
disconnected diagrams in lattice QCD could be used to improve phenomenological PDF extractions. In
particular, we examine the suggestion that the disconnected diagrams have a physical interpretation
and do not contribute to the $\bar{d} - \bar{u}$ asymmetry.

In the following we demonstrate that neither the connected nor the disconnected diagrams in lattice
QCD have a physical meaning on their own.  Each sector contains only a subset of the Wick
contractions generated in performing the Grassmann algebra of the QCD path integral.  Only the
combination of both sectors ensures that the particles propagating are physical.

Indeed, we will show the presence of unphysical states propagating in the valence and sea sectors
and demonstrate how some of the disconnected sea-quark loop sector is used to cancel off unphysical
states propagating in the connected sector.  Two consequences follow from this.  First, any
suggestion that the sum of all the disconnected diagrams must have a physical interpretation is
false.  Part of this contribution is used to remove unphysical hadrons propagating in the connected
sector. Second, the disconnected sea-quark loop contributions remaining no longer have a balance
between $\bar d$ and $\bar u$ contributions.  This latter point admits a nontrivial contribution to
$\bar{d} - \bar{u}$ asymmetry from the disconnected sea-quark loop sector.

In other words, the symmetry required to ensure the disconnected sea-quark loop sector cannot
contribute to the $\bar{d} - \bar{u}$ asymmetry necessarily requires the inclusion of unphysical
baryon contributions, baryons that do not appear in the physical spectrum.

Because these unphysical contributions must be eliminated from the disconnected sea-quark loop
sector through a cancellation with appropriate contributions in the connected sector, the assertion
that the physical $\bar d$ and $\bar u$ disconnected sea-quark loop contributions cancel in
$\bar{d} - \bar{u}$ is incorrect. Upon considering only physically observed baryons in the loop
contributions, we illustrate an important contribution from the disconnected sea-quark loop sector
to the $\bar d - \bar u$ asymmetry.

\section{Quark flow analysis}

We begin our analysis through the consideration of quark flow diagrams included in lattice QCD
calculations of three-point functions of standard proton interpolating fields and a
flavor-diagonal quark-bilinear current.  In performing the Wick contractions, one encounters two
topologically distinct diagrams, one in which the quark flow is connected through the creation
interpolating field to the annihilation interpolating field of the proton, and a second in which
the quark fields of the current are contracted to form a loop.  Such a diagram is commonly referred
to as a quark-flow disconnected diagram.  However, it is connected to the two-point function
through gluon interactions.  The elementary considerations of interpolating fields, the direct and
exchange Wick contractions included in the connected quark-flow diagrams, and the contractions of
the fermion fields in the bilinear current of the three-point function giving rise to quark-flow
disconnected contributions may be reviewed in Ref.~\cite{Leinweber:1995ie}.

In drawing the diagrams associated with the disconnected sea-quark loop contribution, one can
consider flavor-singlet constructions where the quark-antiquark pair of the loop annihilates to
gluons, and flavor-octet contributions where the loop pairs with one of the quark-flow lines of
the proton interpolating fields to form flavor-octet mesons.  As the flavor-singlet mesons have an
equal balance of $\bar u u$ and $\bar d d$ components in the mass-symmetric limit neglecting
electromagnetic interactions, these contributions do not generate a $\bar d - \bar u$ asymmetry and
we do not consider them further.

\begin{figure*}[t]
\includegraphics[width=0.7\textwidth]{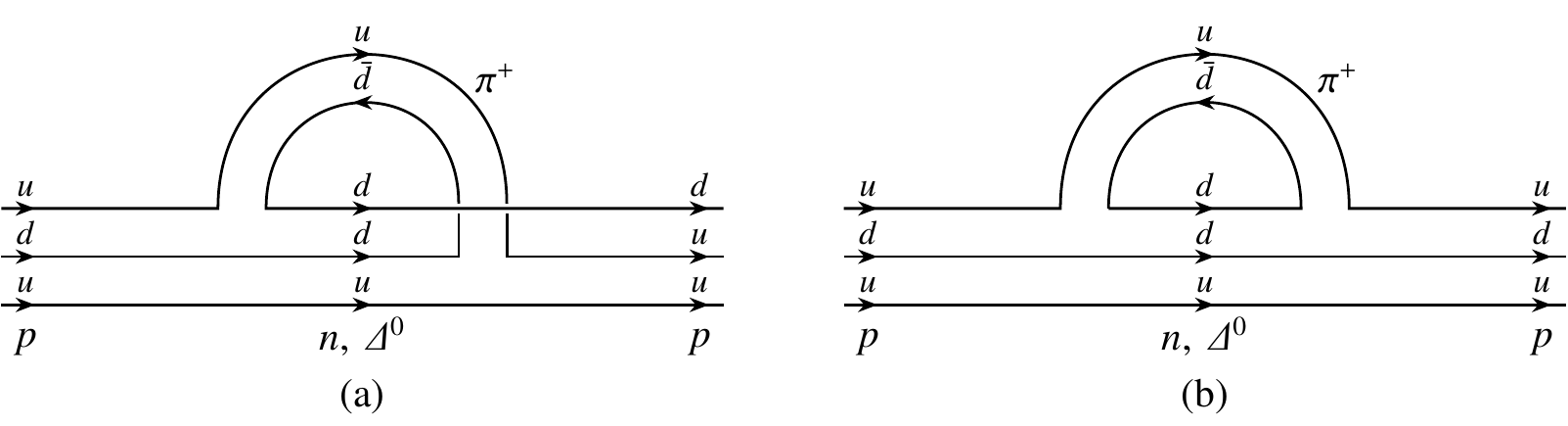}
\includegraphics[width=0.7\textwidth]{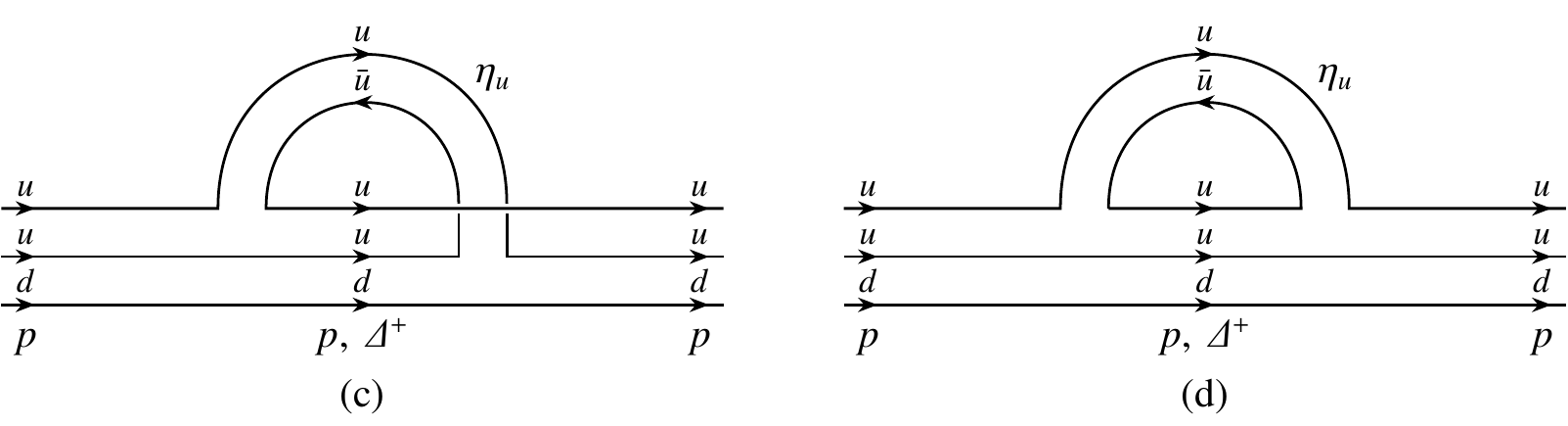}
\includegraphics[width=0.7\textwidth]{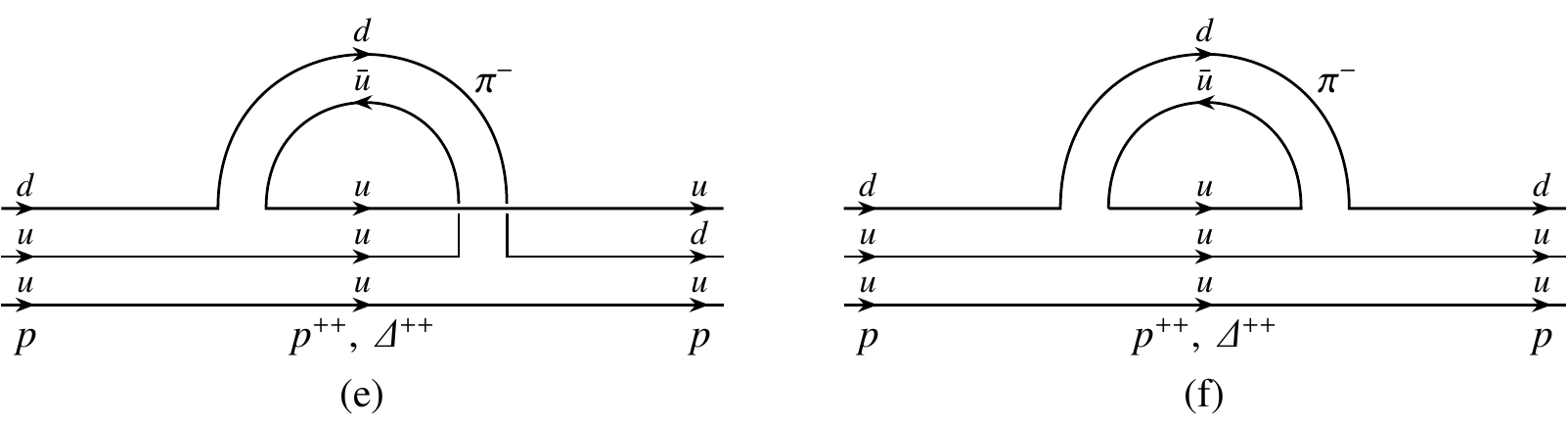}
\includegraphics[width=0.7\textwidth]{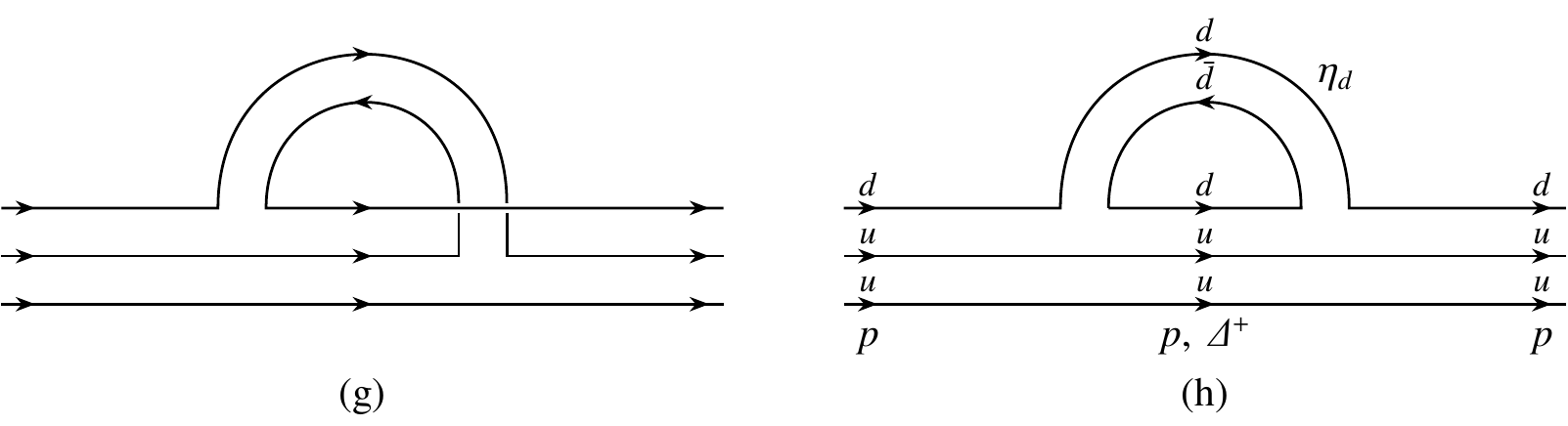}
\vspace{12pt}
\caption{Quark flow diagrams describing the one-loop meson dressings contributing to the $\bar d -
  \bar u$ asymmetry of the proton. The connected flows are listed in the left column as diagrams
  (a), (c), (e) and (g). The corresponding flows incorporating a disconnected sea-quark loop
  contribution appear in the right-hand column as diagrams (b), (d), (f), and (h).  Diagram (g)
  contains no quark labels and serves to illustrate there is no connected contribution to the $d
  \bar d$ meson component. We focus on contributions where the external current acts on the
  antiquark loop. }
\label{fig:quarkFlow}
\end{figure*}

Figure \ref{fig:quarkFlow} illustrates the quark-flow diagrams having overlap with meson dressings
of the proton contributing to the $\bar d - \bar u$ asymmetry of the proton.  The eight diagrams
are associated with two choices for the light quark propagating in the meson dressing, times two
choices for the antiquark, times two topologically different quark flows giving rise to the meson
dressing.  Connected Wick contractions, including both the direct and exchange contractions, are
listed in the left column and the corresponding quark-bilinear current contractions generating a
disconnected sea-quark loop contribution appear in the right-hand column.

There are three connected flows to consider as the $d$ quark can be placed on each of the quark
flow lines.  With the intermediate-state quark-flows identified, the corresponding diagram
incorporating a disconnected sea-quark loop is illustrated.  In obtaining the physical
contribution, both the connected and disconnected sea-quark loop contributions are summed.  The
final row, illustrating the $d \bar d$ component of the meson dressings emphasizes that there is no
connected contribution to this component.  One would need two $d$ quarks in the valence sector of
the proton.  Thus, in this case, only the disconnected sea-quark loop contributes.

We now turn our attention to the symmetries manifest in lattice QCD calculations.  These
calculations are not restricted to specific choices for the mesons and baryons participating in the
intermediate state.  Rather, the lattice correlation functions have overlap with all baryons and
mesons associated with the quantum numbers specified by the quark flow.  Thus, in this discussion
of the symmetries of lattice QCD correlation functions, we will ignore the hadron labels on the
intermediate states of Fig.~ \ref{fig:quarkFlow}.

The symmetry of the $\bar u$ and $\bar d$ contributions is manifest in the right-hand column of
Fig.~\ref{fig:quarkFlow}.  When a $u$ valence quark is contributing to a meson dressing, both the
$\bar d$ and $\bar u$ sea quarks contribute in diagrams (b) and (d) respectively.  For equal-mass
$u$ and $d$ flavors, and neglecting electromagnetic effects, these sea-quark loop contributions
are equal.  While the gluonic interactions are sensitive to the mass of the quarks, they do not
differentiate flavor.  Similarly, when a $d$ valence quark is contributing to a meson dressing,
both the $\bar u$ and $\bar d$ sea quarks contribute in diagrams (f) and (h) respectively, again
maintaining the symmetry.  The naive conclusion is that this symmetry prevents a contribution to
the $\bar d - \bar u$ asymmetry.  However there is a subtlety that has been overlooked.

If one considers the connected and disconnected diagrams on their own, without taking the sum, one
must expand the list of baryons and mesons having overlap with the quark flows to include
unphysical states that do not appear in the physical spectrum.  Indeed, the Pauli exclusion
principle is enforced by summing both of the aforementioned Wick contractions leading to connected
and disconnected quark flows.  On their own, a connected or disconnected graph will receive
contributions from baryons that do not satisfy the Pauli exclusion principle.  Fortunately, the
symmetries highlighted in the previous paragraph can be used to determine the properties of these
unphysical states.

At this point, it is helpful to become more specific on the hadrons contributing to the quark flows
of Fig.~\ref{fig:quarkFlow}.  Thus we turn our attention to light pseudoscalar meson dressings
having intermediate nucleon or $\Delta$-baryon intermediate states.
These dressings provide the most important contributions to the $\bar d - \bar u$ asymmetry.
Dressings with intermediate baryons degenerate with the nucleon, not only isolate the leading
nonanalytic contribution in chiral perturbation
theory~\cite{Thomas:2000ny,Ji:2013bca,Chen:2001eg,Arndt:2001ye}, but also serve to expose the
presence of unphysical baryon contributions.  The consideration of $\Delta$ intermediate states
serves to further illustrate the necessity of an unphysical baryon degenerate with the nucleon.

We commence by focusing on dressings with intermediate baryons degenerate with the nucleon.
Diagrams (a) and (b) of Fig.~\ref{fig:quarkFlow} sum to provide the physical $n\, \pi^+$ dressing
of the proton.  In diagrams (c) and (d), we are introduced to the $\eta_u$ meson, a meson composed
of a $u \bar u$ pair.  Similarly, the $d \bar d$ meson in diagram (h) is labeled as $\eta_d$.  The
masses of these neutral pseudoscalar mesons can be inferred from the symmetries manifest at the
level of lattice QCD correlation functions.

Because the sea-quark loop contributions from diagrams (b) and (d) are equal, we conclude that when
the $\bar d u$ meson of diagram (b) is a $\pi^+$, the $\eta_u$ meson of diagram (d) has a mass
degenerate with the pion.  Similarly, because the sea-quark loop contributions from diagrams (f)
and (h) are equal, when the $\bar u d$ meson of diagram (f) is a $\pi^-$, the $\eta_d$ meson of
diagram (h) has a mass degenerate with the pion.

The axial couplings of these mesons are provided in a meson basis in which the diagonal entries of
the nonet pseudoscalar meson matrix are $\eta_u$, $\eta_d$, and $\eta_s$.  In other words, the
square of the axial couplings of the $\eta_u$ and $\eta_d$ mesons is associated with a linear
combination of $\pi^0$, $\eta$, and $\eta'$ mesons in the proportion $3:1:2$ respectively for
standard SU(3) symmetry.  We emphasize this is simply a change of basis.  The masses of the mesons
are governed by the symmetries encountered in lattice QCD calculations.

\begin{table*}[t]
\caption{The leading contribution for each diagram of Fig.~\ref{fig:quarkFlow} from SU(3)
  partially quenched chiral perturbation theory.  Here we use the normalization of
  Ref.~\cite{Leinweber:2002qb} that sums diagrams (a) and (b) to give the physical $n\,\pi^+$
  dressing of the proton equal to $2\,(D+F)^2 = 2\, g_A^2$.  Numerical values correspond to $D =
  0.77$ and $F=0.50$.  The right-hand column entitled Relative Contribution describes the
  percentage split of the strength between the connected and disconnected contributions in each row
  of Fig.~\ref{fig:quarkFlow}.}
\label{tab:coefficients}
\centering
\begin{tabular}{clcc}
\noalign{\smallskip}
\hline
\noalign{\smallskip}
        &                                 &            &Relative \\
Diagram &Contribution                     &Value       &Contribution (\%) \\
\noalign{\smallskip}
\hline
\noalign{\smallskip}
  (a)   &$2\,D^2 / 3 + 4\,D\,F - 2\,F^2$  &$ 1.44$     &$ 44.5$ \\
  (b)   &$(D + 3\,F)^2 / 3 + (D - F)^2 $  &$ 1.79$     &$ 55.5$ \\
  (c)   &$4\,(F^2 - D^2 / 3)           $  &$ 0.21$     &$ 10.5$ \\
  (d)   &$(D + 3\,F)^2 / 3 + (D - F)^2 $  &$ 1.79$     &$ 89.5$ \\
  (e)   &$-2\,(D - F)^2                $  &$-0.15$     &$ 50.0$ \\
  (f)   &$ 2\,(D - F)^2                $  &$ 0.15$     &$ 50.0$ \\
  (h)   &$ 2\,(D - F)^2                $  &$ 0.15$     &$100.0$ \\
\noalign{\smallskip}
\hline
\end{tabular}
\end{table*}

The low-lying baryons associated with the quark flows are also labeled in
Fig.~\ref{fig:quarkFlow}.
It is here that one encounters an unphysical baryon labeled ``$p^{++}$'' in diagrams (e) and (f).
While the charge of $+2$ is manifest in the presence of a $\pi^-$ and the necessity of charge
conservation, the label of a $p$ emphasizing the propagation of a baryon degenerate with the proton
requires further explanation.  

Because QCD interactions are flavor blind when $m_u = m_d$, diagrams (f) and (h) make equal
contributions.  It is clear that diagram (h) has overlap with the normal proton propagating in the
intermediate state.  Indeed diagrams (c), (d) and (h) sum to provide the physical $p\, \pi^0$
dressing of the proton.  Thus, through the equivalence of diagrams (f) and (h) established by the
symmetries of lattice QCD calculations, one concludes that the mass of the $p^{++}$ is also that of
the proton.  The mass of the intermediate proton in diagram (h) excludes the identification of the
$p^{++}$ as the physical $\Delta^{++}$.

Because the Pauli exclusion principle does not allow octet baryons with charge $+2$, we must look
to the other connected Wick contractions to cancel the unphysical contribution of the
sea-quark loop in diagram (f).  This is the role of diagram (e) and thus we conclude that the
contribution of diagram (e) is equal and opposite to that of diagram (f).  In other words, if we
were to restrict our diagrams to those having physical intermediate baryons, we would need to
eliminate diagrams (e) and (f) in the third row of Fig.~\ref{fig:quarkFlow}.  With the loss of
diagram (f), the symmetry of the disconnected sea-quark loop contributions is lost.
Flavor-blindness in QCD interactions equates the contributions of diagrams (b) and (d), leaving
the contribution from diagram (h) to break the symmetry.  Thus, we conclude that disconnected
sea-quark loops do in fact make physical contributions to the $\bar d - \bar u$ asymmetry in the
proton.

We now turn our attention to pion-loop dressings with $\Delta$-baryon intermediate states.  A
survey of these baryon labels in Fig.~\ref{fig:quarkFlow} reveals all of the required charge
assignments are associated with physical $\Delta$ states.  Indeed, the problem encountered in
considering diagrams (e) and (f) where a $p^{++}$ baryon degenerate with the proton was required,
now presents no problem.  This time, diagram (h) has overlap with a $\Delta^+$ propagating
in the intermediate state. Now the equivalence of diagrams (f) and (h) established by the
symmetries of lattice QCD calculations, demands that the mass of the intermediate baryon
propagating in diagram (f) is degenerate with the $\Delta^+$.  Noting the $\Delta$ baryons are
degenerate in the mass-symmetric limit when electromagnetic effects are neglected, the $\Delta^{++}$
baryon of diagram (f) satisfies this condition.

In summary, all the intermediate $\Delta$-baryon states can be associated with physical $\Delta$
states.  Cancellation between diagrams (e) and (f) is not required for $\Delta$-baryon intermediate
states.  In this case, diagram (f) remains available to complete the symmetry of $\bar d$ and $\bar
u$ disconnected sea-quark loop contributions and the disconnected sea-quark loop sector does not
generate a contribution to the $\bar d - \bar u$ asymmetry of the nucleon when the intermediate
propagating baryon is a $\Delta$ baryon.  As such, we will not consider $\Delta$ baryons further.

\section{Partially quenched chiral perturbation theory}

Partially quenched chiral perturbation theory for baryons
\cite{Labrenz:1996jy,Savage:2001dy,Chen:2001yi,Leinweber:2002qb,Hall:2015cua} provides a mechanism
%
%
%
%
%
for understanding the relative contributions of the connected and disconnected quark-flow diagrams.
While the label ``partially quenched'' has an historical origin, the results presented here pertain
to that of physical QCD. Here we draw on the results in Tables I and II of
Ref.~\cite{Leinweber:2002qb} using the diagrammatic approach for obtaining the meson-baryon
couplings in terms of the familiar SU(3)-flavor axial couplings, $F$ and $D$.  There the physical
basis was used for the axial couplings and the $\eta_u$ and $\eta_d$ couplings are obtained by
summing over the $\pi^0$, $\eta$, and $\eta'$ entries.

Our focus is now on the meson-loop dressings of Fig.~\ref{fig:quarkFlow} with intermediate baryons
degenerate with the proton.  
%
%
These dressings provide the leading nonanalytic contributions to the $\bar d - \bar u$ asymmetry
and we refer to them succinctly as the leading contributions in the following.  

Diagrams (a) and (b) of Fig.~\ref{fig:quarkFlow} sum to generate the full physical $\pi^+$ dressing
of the proton, The disconnected sea-quark loop coefficients are provided in Table
II of Ref.~\cite{Leinweber:2002qb} and the coefficients of the connected diagrams are obtained by
subtracting the disconnected contributions from the full contributions provided in
Table I of Ref.~\cite{Leinweber:2002qb}.

Working with the normalization of Ref.~\cite{Leinweber:2002qb} where the contribution of the $n\,
\pi^+$ dressing of the proton is proportional to $2\,(D+F)^2 = 2\, g_A^2$, the contribution of each
diagram in Fig.~\ref{fig:quarkFlow} generating the leading contribution is summarized in
Table~\ref{tab:coefficients}.  Here the numerical values correspond to $D = 0.77$ and $F=0.50$,
such that $g_A=1.27$.

Note how the symmetries discussed in the previous section in the context of lattice QCD quark-flow
diagrams are manifest in Table~\ref{tab:coefficients} for pseudoscalar dressings of the proton.
The equivalence of diagrams (b) and (d) is reflected in the equivalence of the axial couplings for
these diagrams.  The equivalence of diagrams (f) and (h) is similarly reflected in equal axial
couplings for those diagrams.

With regard to the $\bar d - \bar u$ asymmetry of the nucleon, we recall that diagram (f) is
combined with (e) to eliminate the unphysical propagation of the $p^{++}$.  While the contributions
of diagrams (b) and (d) maintain the symmetry of $\bar d$ and $\bar u$ contributions in the
disconnected sea-quark loop contributions, diagram (h) stands alone.  As a result, there is an
enhancement of $\bar d$ in the disconnected sea-quark loop sector coming from diagram (h).  The
symmetry is broken as unphysical baryons are removed from the intermediate states.  Thus, the
assumption of equal $\bar d$ and $\bar u$ in the disconnected sector of the sea is wrong in
principle.

The axial couplings of Table I allow one to quantify the relative contributions of the connected
and disconnected sectors to $\bar d - \bar u$.  Commencing with the connected sector, the third
column of Table I provides the couplings for diagrams (a) and (c).  Their difference leaves a net
effect of $1.44 - 0.21 = 1.23$ favoring the $\bar d$ contribution.  Now the disconnected
sea-quark loop contribution of diagram (h) further enhances the $\bar d$ excess, raising it from
$1.23 \to 1.23 + 0.15 = 1.38$.  Thus a 12\% enhancement of the pion-nucleon contribution to $\bar d
- \bar u$ in the proton has its origin in the disconnected sea-quark loop sector.

\section{Conclusions}

We have analyzed the role of connected and disconnected quark-flow diagrams in generating a $\bar d
- \bar u$ asymmetry in the sea of the proton.  The formalisms of lattice QCD and partially quenched
chiral perturbation theory have been considered.  It has been shown that the naive assertion of the
equality of the $\bar d$ and $\bar u$ contributions from disconnected sea-quark loops relies on
contributions from an unphysical baryon.  There is an unphysical contribution to $\bar u$ arising
from the process involving an intermediate $\pi^-$ and a charge $2^{++}$ baryon degenerate with the
proton, illustrated in Fig.~1(f). In a full QCD calculation this unphysical contribution is
precisely cancelled by the connected contribution to $\bar u$ shown in Fig.~1(e). Once the
unphysical contribution associated with Fig.~1(f) is removed, one finds an enhancement of
approximately 12\% in the $\bar d - \bar u$ asymmetry arising from pion-baryon-octet components of
the proton wave function.

\section*{Acknowledgements}

We are pleased to acknowledge helpful discussions with Anthony Williams and Ross Young. This
research was undertaken with the assistance of resources from the National Computational
Infrastructure (NCI), provided through the National Computational Merit Allocation Scheme.
This research was supported by the Australian Research Council through ARC Discovery Project Grants
DP190102215 and DP210103706 (D.B.L.).

\bibliographystyle{utphys}
\bibliography{refs}

\providecommand{\href}[2]{#2}\begingroup\raggedright\begin{thebibliography}{10}

\bibitem{Gross:2022hyw}
F.~Gross {\em et~al.}, ``{50 Years of Quantum Chromodynamics},''
  \href{http://dx.doi.org/10.1140/epjc/s10052-023-11949-2}{{\em Eur. Phys. J.
  C} {\bfseries 83} (2023) 1125},
  \href{http://arxiv.org/abs/2212.11107}{{\ttfamily arXiv:2212.11107
  [hep-ph]}}.

\bibitem{Thomas:1983fh}
A.~W. Thomas, ``{A Limit on the Pionic Component of the Nucleon Through SU(3)
  Flavor Breaking in the Sea},''
  \href{http://dx.doi.org/10.1016/0370-2693(83)90026-6}{{\em Phys. Lett. B}
  {\bfseries 126} (1983) 97--100}.

\bibitem{Ito:1980ev}
A.~S. Ito {\em et~al.}, ``{Measurement of the Continuum of Dimuons Produced in
  High-Energy Proton - Nucleus Collisions},''
  \href{http://dx.doi.org/10.1103/PhysRevD.23.604}{{\em Phys. Rev. D}
  {\bfseries 23} (1981) 604--633}.

\bibitem{NewMuonNMC:1990xyw}
{\bfseries New Muon (NMC)} Collaboration, D.~Allasia {\em et~al.},
  ``{Measurement of the neutron and the proton $F_2$ structure function
  ratio},'' \href{http://dx.doi.org/10.1016/0370-2693(90)91270-L}{{\em Phys.
  Lett. B} {\bfseries 249} (1990) 366--372}.

\bibitem{Gottfried:1967kk}
K.~Gottfried, ``{Sum rule for high-energy electron - proton scattering},''
  \href{http://dx.doi.org/10.1103/PhysRevLett.18.1174}{{\em Phys. Rev. Lett.}
  {\bfseries 18} (1967) 1174}.

\bibitem{Gottfried:1986hc}
K.~Gottfried and V.~F. Weisskopf, {\em {Concepts of Particle Physics. Vol. 2}}.
\newblock Oxford University Press, 1986.

\bibitem{SeaQuest:2021zxb}
{\bfseries SeaQuest} Collaboration, J.~Dove {\em et~al.}, ``{The asymmetry of
  antimatter in the proton},''
  \href{http://dx.doi.org/10.1038/s41586-022-04707-z}{{\em Nature} {\bfseries
  590} no.~7847, (2021) 561--565},
  \href{http://arxiv.org/abs/2103.04024}{{\ttfamily arXiv:2103.04024
  [hep-ph]}}. [Erratum: Nature 604, E26 (2022)].

\bibitem{NuSea:2001idv}
{\bfseries NuSea} Collaboration, R.~S. Towell {\em et~al.}, ``{Improved
  measurement of the $\bar d / \bar u$ asymmetry in the nucleon sea},''
  \href{http://dx.doi.org/10.1103/PhysRevD.64.052002}{{\em Phys. Rev. D}
  {\bfseries 64} (2001) 052002},
  \href{http://arxiv.org/abs/hep-ex/0103030}{{\ttfamily arXiv:hep-ex/0103030}}.

\bibitem{NewMuon:1993oys}
{\bfseries New Muon} Collaboration, M.~Arneodo {\em et~al.}, ``{A Reevaluation
  of the Gottfried sum},'' \href{http://dx.doi.org/10.1103/PhysRevD.50.R1}{{\em
  Phys. Rev. D} {\bfseries 50} (1994) R1--R3}.

\bibitem{Thomas:2000ny}
A.~W. Thomas, W.~Melnitchouk, and F.~M. Steffens, ``{Dynamical symmetry
  breaking in the sea of the nucleon},''
  \href{http://dx.doi.org/10.1103/PhysRevLett.85.2892}{{\em Phys. Rev. Lett.}
  {\bfseries 85} (2000) 2892--2894},
  \href{http://arxiv.org/abs/hep-ph/0005043}{{\ttfamily arXiv:hep-ph/0005043}}.

\bibitem{He:2022fne}
F.~He, C.-R. Ji, W.~Melnitchouk, Y.~Salamu, A.~W. Thomas, P.~Wang, and X.~G.
  Wang, ``{Helicity-dependent distribution of strange quarks in the proton from
  nonlocal chiral effective theory},''
  \href{http://dx.doi.org/10.1103/PhysRevD.105.094007}{{\em Phys. Rev. D}
  {\bfseries 105} no.~9, (2022) 094007},
  \href{http://arxiv.org/abs/2203.06628}{{\ttfamily arXiv:2203.06628
  [hep-ph]}}.

\bibitem{Kretzer:2003wy}
S.~Kretzer, F.~Olness, J.~Pumplin, D.~Stump, W.-K. Tung, and M.~H. Reno, ``{The
  Parton structure of the nucleon and precision determination of the Weinberg
  angle in neutrino scattering},''
  \href{http://dx.doi.org/10.1103/PhysRevLett.93.041802}{{\em Phys. Rev. Lett.}
  {\bfseries 93} (2004) 041802},
  \href{http://arxiv.org/abs/hep-ph/0312322}{{\ttfamily arXiv:hep-ph/0312322}}.

\bibitem{Salamu:2019dok}
Y.~Salamu, C.-R. Ji, W.~Melnitchouk, A.~W. Thomas, P.~Wang, and X.~G. Wang,
  ``{Parton distributions from nonlocal chiral SU(3) effective theory: Flavor
  asymmetries},'' \href{http://dx.doi.org/10.1103/PhysRevD.100.094026}{{\em
  Phys. Rev. D} {\bfseries 100} no.~9, (2019) 094026},
  \href{http://arxiv.org/abs/1907.08551}{{\ttfamily arXiv:1907.08551
  [hep-ph]}}.

\bibitem{Wang:2016ndh}
X.~G. Wang, C.-R. Ji, W.~Melnitchouk, Y.~Salamu, A.~W. Thomas, and P.~Wang,
  ``{Strange quark asymmetry in the proton in chiral effective theory},''
  \href{http://dx.doi.org/10.1103/PhysRevD.94.094035}{{\em Phys. Rev. D}
  {\bfseries 94} no.~9, (2016) 094035},
  \href{http://arxiv.org/abs/1610.03333}{{\ttfamily arXiv:1610.03333
  [hep-ph]}}.

\bibitem{Burkardt:2012hk}
M.~Burkardt, K.~S. Hendricks, C.-R. Ji, W.~Melnitchouk, and A.~W. Thomas,
  ``{Pion momentum distributions in the nucleon in chiral effective theory},''
  \href{http://dx.doi.org/10.1103/PhysRevD.87.056009}{{\em Phys. Rev. D}
  {\bfseries 87} no.~5, (2013) 056009},
  \href{http://arxiv.org/abs/1211.5853}{{\ttfamily arXiv:1211.5853 [hep-ph]}}.

\bibitem{Alberg:2017ijg}
M.~Alberg and G.~A. Miller, ``{Chiral Light Front Perturbation Theory and the
  Flavor Dependence of the Light-Quark Nucleon Sea},''
  \href{http://dx.doi.org/10.1103/PhysRevC.100.035205}{{\em Phys. Rev. C}
  {\bfseries 100} no.~3, (2019) 035205},
  \href{http://arxiv.org/abs/1712.05814}{{\ttfamily arXiv:1712.05814
  [nucl-th]}}.

\bibitem{Diakonov:1996sr}
D.~Diakonov, V.~Petrov, P.~Pobylitsa, M.~V. Polyakov, and C.~Weiss, ``{Nucleon
  parton distributions at low normalization point in the large $N_c$ limit},''
  \href{http://dx.doi.org/10.1016/S0550-3213(96)00486-5}{{\em Nucl. Phys. B}
  {\bfseries 480} (1996) 341--380},
  \href{http://arxiv.org/abs/hep-ph/9606314}{{\ttfamily arXiv:hep-ph/9606314}}.

\bibitem{Dressler:1999zv}
B.~Dressler, K.~Goeke, M.~V. Polyakov, P.~Schweitzer, M.~Strikman, and
  C.~Weiss, ``{Polarized anti-quark flavor asymmetry in Drell-Yan pair
  production},'' \href{http://dx.doi.org/10.1007/s100520100567}{{\em Eur. Phys.
  J. C} {\bfseries 18} (2001) 719--722},
  \href{http://arxiv.org/abs/hep-ph/9910464}{{\ttfamily arXiv:hep-ph/9910464}}.

\bibitem{Melnitchouk:1991ui}
W.~Melnitchouk, A.~W. Thomas, and A.~I. Signal, ``{Gottfried sum rule and the
  shape of $F_2^p - F_2^n$},'' \href{http://dx.doi.org/10.1007/BF01284484}{{\em
  Z. Phys. A} {\bfseries 340} (1991) 85--92}.

\bibitem{Henley:1990kw}
E.~M. Henley and G.~A. Miller, ``{Excess of $\bar d$ over $\bar u$ in the
  proton sea quark distribution},''
  \href{http://dx.doi.org/10.1016/0370-2693(90)90735-O}{{\em Phys. Lett. B}
  {\bfseries 251} (1990) 453--454}.

\bibitem{Signal:1987gz}
A.~I. Signal and A.~W. Thomas, ``{Possible Strength of the Nonperturbative
  Strange Sea of the Nucleon},''
  \href{http://dx.doi.org/10.1016/0370-2693(87)91348-7}{{\em Phys. Lett. B}
  {\bfseries 191} (1987) 205}.

\bibitem{NNPDF:2021njg}
{\bfseries NNPDF} Collaboration, R.~D. Ball {\em et~al.}, ``{The path to proton
  structure at 1\% accuracy},''
  \href{http://dx.doi.org/10.1140/epjc/s10052-022-10328-7}{{\em Eur. Phys. J.
  C} {\bfseries 82} no.~5, (2022) 428},
  \href{http://arxiv.org/abs/2109.02653}{{\ttfamily arXiv:2109.02653
  [hep-ph]}}.

\bibitem{AbdulKhalek:2022hcn}
R.~Abdul~Khalek {\em et~al.}, ``{Snowmass 2021 White Paper: Electron Ion
  Collider for High Energy Physics},'' in {\em {2022 Snowmass Summer Study}}.
\newblock 3, 2022.
\newblock \href{http://arxiv.org/abs/2203.13199}{{\ttfamily arXiv:2203.13199
  [hep-ph]}}.

\bibitem{Thomas:2021lub}
A.~W. Thomas, X.~G. Wang, and A.~G. Williams, ``{Constraints on the dark photon
  from deep inelastic scattering},''
  \href{http://dx.doi.org/10.1103/PhysRevD.105.L031901}{{\em Phys. Rev. D}
  {\bfseries 105} no.~3, (2022) L031901},
  \href{http://arxiv.org/abs/2111.05664}{{\ttfamily arXiv:2111.05664
  [hep-ph]}}.

\bibitem{NNPDF:2017mvq}
{\bfseries NNPDF} Collaboration, R.~D. Ball {\em et~al.}, ``{Parton
  distributions from high-precision collider data},''
  \href{http://dx.doi.org/10.1140/epjc/s10052-017-5199-5}{{\em Eur. Phys. J. C}
  {\bfseries 77} no.~10, (2017) 663},
  \href{http://arxiv.org/abs/1706.00428}{{\ttfamily arXiv:1706.00428
  [hep-ph]}}.

\bibitem{Zheng:2021hcf}
X.~Zheng, J.~Erler, Q.~Liu, and H.~Spiesberger, ``{Accessing weak
  neutral-current coupling $g_{AA}^{eq}$ using positron and electron beams at
  Jefferson Lab},''
  \href{http://dx.doi.org/10.1140/epja/s10050-021-00490-z}{{\em Eur. Phys. J.
  A} {\bfseries 57} no.~5, (2021) 173},
  \href{http://arxiv.org/abs/2103.12555}{{\ttfamily arXiv:2103.12555
  [nucl-ex]}}.

\bibitem{Bentz:2009yy}
W.~Bentz, I.~C. Cloet, J.~T. Londergan, and A.~W. Thomas, ``{Reassessment of
  the NuTeV determination of the weak mixing angle},''
  \href{http://dx.doi.org/10.1016/j.physletb.2010.09.001}{{\em Phys. Lett. B}
  {\bfseries 693} (2010) 462--466},
  \href{http://arxiv.org/abs/0908.3198}{{\ttfamily arXiv:0908.3198 [nucl-th]}}.

\bibitem{Martin:2009iq}
A.~D. Martin, W.~J. Stirling, R.~S. Thorne, and G.~Watt, ``{Parton
  distributions for the LHC},''
  \href{http://dx.doi.org/10.1140/epjc/s10052-009-1072-5}{{\em Eur. Phys. J. C}
  {\bfseries 63} (2009) 189--285},
  \href{http://arxiv.org/abs/0901.0002}{{\ttfamily arXiv:0901.0002 [hep-ph]}}.

\bibitem{Detmold:2020snb}
{\bfseries NPLQCD} Collaboration, W.~Detmold, M.~Illa, D.~J. Murphy, P.~Oare,
  K.~Orginos, P.~E. Shanahan, M.~L. Wagman, and F.~Winter, ``{Lattice QCD
  Constraints on the Parton Distribution Functions of $^3$He},''
  \href{http://dx.doi.org/10.1103/PhysRevLett.126.202001}{{\em Phys. Rev.
  Lett.} {\bfseries 126} no.~20, (2021) 202001},
  \href{http://arxiv.org/abs/2009.05522}{{\ttfamily arXiv:2009.05522
  [hep-lat]}}.

\bibitem{JeffersonLabAngularMomentumJAM:2022aix}
{\bfseries Jefferson Lab Angular Momentum (JAM), HadStruc} Collaboration, P.~C.
  Barry {\em et~al.}, ``{Complementarity of experimental and lattice QCD data
  on pion parton distributions},''
  \href{http://dx.doi.org/10.1103/PhysRevD.105.114051}{{\em Phys. Rev. D}
  {\bfseries 105} no.~11, (2022) 114051},
  \href{http://arxiv.org/abs/2204.00543}{{\ttfamily arXiv:2204.00543
  [hep-ph]}}.

\bibitem{Hou:2022ajg}
T.-J. Hou, M.~Yan, J.~Liang, K.-F. Liu, and C.~P. Yuan, ``{Connected and
  disconnected sea partons from the CT18 parametrization of PDFs},''
  \href{http://dx.doi.org/10.1103/PhysRevD.106.096008}{{\em Phys. Rev. D}
  {\bfseries 106} no.~9, (2022) 096008},
  \href{http://arxiv.org/abs/2206.02431}{{\ttfamily arXiv:2206.02431
  [hep-ph]}}.

\bibitem{Liu:2012ch}
K.-F. Liu, W.-C. Chang, H.-Y. Cheng, and J.-C. Peng, ``{Connected-Sea
  Partons},'' \href{http://dx.doi.org/10.1103/PhysRevLett.109.252002}{{\em
  Phys. Rev. Lett.} {\bfseries 109} (2012) 252002},
  \href{http://arxiv.org/abs/1206.4339}{{\ttfamily arXiv:1206.4339 [hep-ph]}}.

\bibitem{Zhang:2001gna}
Y.-j. Zhang, B.~Zhang, and B.-Q. Ma, ``{Detailed balance and sea quark flavor
  asymmetry of proton},''
  \href{http://dx.doi.org/10.1016/S0370-2693(01)01266-7}{{\em Phys. Lett. B}
  {\bfseries 523} (2001) 260--264},
  \href{http://arxiv.org/abs/hep-ph/0106074}{{\ttfamily arXiv:hep-ph/0106074}}.

\bibitem{Liu:1993cv}
K.-F. Liu and S.-J. Dong, ``{Origin of difference between $\bar d$ and $\bar u$
  partons in the nucleon},''
  \href{http://dx.doi.org/10.1103/PhysRevLett.72.1790}{{\em Phys. Rev. Lett.}
  {\bfseries 72} (1994) 1790--1793},
  \href{http://arxiv.org/abs/hep-ph/9306299}{{\ttfamily arXiv:hep-ph/9306299}}.

\bibitem{Liu:1999ak}
K.-F. Liu, ``{Parton degrees of freedom from the path integral formalism},''
  \href{http://dx.doi.org/10.1103/PhysRevD.62.074501}{{\em Phys. Rev. D}
  {\bfseries 62} (2000) 074501},
  \href{http://arxiv.org/abs/hep-ph/9910306}{{\ttfamily arXiv:hep-ph/9910306}}.

\bibitem{Leinweber:1995ie}
D.~B. Leinweber, ``{QCD equalities for baryon current matrix elements},''
  \href{http://dx.doi.org/10.1103/PhysRevD.53.5115}{{\em Phys. Rev.} {\bfseries
  D53} (1996) 5115--5124},
\href{http://arxiv.org/abs/hep-ph/9512319}{{\ttfamily arXiv:hep-ph/9512319
  [hep-ph]}}.

\bibitem{Ji:2013bca}
C.-R. Ji, W.~Melnitchouk, and A.~W. Thomas, ``{Anatomy of relativistic pion
  loop corrections to the electromagnetic nucleon coupling},''
  \href{http://dx.doi.org/10.1103/PhysRevD.88.076005}{{\em Phys. Rev. D}
  {\bfseries 88} no.~7, (2013) 076005},
  \href{http://arxiv.org/abs/1306.6073}{{\ttfamily arXiv:1306.6073 [hep-ph]}}.

\bibitem{Chen:2001eg}
J.-W. Chen and X.-d. Ji, ``{Is the Sullivan process compatible with QCD chiral
  dynamics?},'' \href{http://dx.doi.org/10.1016/S0370-2693(01)01337-5}{{\em
  Phys. Lett. B} {\bfseries 523} (2001) 107--110},
  \href{http://arxiv.org/abs/hep-ph/0105197}{{\ttfamily arXiv:hep-ph/0105197}}.

\bibitem{Arndt:2001ye}
D.~Arndt and M.~J. Savage, ``{Chiral corrections to matrix elements of twist-2
  operators},'' \href{http://dx.doi.org/10.1016/S0375-9474(01)01223-4}{{\em
  Nucl. Phys. A} {\bfseries 697} (2002) 429--439},
  \href{http://arxiv.org/abs/nucl-th/0105045}{{\ttfamily
  arXiv:nucl-th/0105045}}.

\bibitem{Leinweber:2002qb}
D.~B. Leinweber, ``{Quark contributions to baryon magnetic moments in full,
  quenched and partially quenched QCD},''
  \href{http://dx.doi.org/10.1103/PhysRevD.69.014005}{{\em Phys. Rev.}
  {\bfseries D69} (2004) 014005},
\href{http://arxiv.org/abs/hep-lat/0211017}{{\ttfamily arXiv:hep-lat/0211017
  [hep-lat]}}.

\bibitem{Labrenz:1996jy}
J.~N. Labrenz and S.~R. Sharpe, ``{Quenched chiral perturbation theory for
  baryons},'' \href{http://dx.doi.org/10.1103/PhysRevD.54.4595}{{\em Phys.
  Rev.} {\bfseries D54} (1996) 4595--4608},
\href{http://arxiv.org/abs/hep-lat/9605034}{{\ttfamily arXiv:hep-lat/9605034
  [hep-lat]}}.

\bibitem{Savage:2001dy}
M.~J. Savage, ``{The Magnetic moments of the octet baryons in quenched chiral
  perturbation theory},''
  \href{http://dx.doi.org/10.1016/S0375-9474(01)01315-X}{{\em Nucl. Phys.}
  {\bfseries A700} (2002) 359--376},
\href{http://arxiv.org/abs/nucl-th/0107038}{{\ttfamily arXiv:nucl-th/0107038
  [nucl-th]}}.

\bibitem{Chen:2001yi}
J.-W. Chen and M.~J. Savage, ``{Baryons in partially quenched chiral
  perturbation theory},''
  \href{http://dx.doi.org/10.1103/PhysRevD.65.094001}{{\em Phys. Rev.}
  {\bfseries D65} (2002) 094001},
\href{http://arxiv.org/abs/hep-lat/0111050}{{\ttfamily arXiv:hep-lat/0111050
  [hep-lat]}}.

\bibitem{Hall:2015cua}
J.~M.~M. Hall and D.~B. Leinweber, ``{Flavor-singlet baryons in the graded
  symmetry approach to partially quenched QCD},''
  \href{http://dx.doi.org/10.1103/PhysRevD.94.094004}{{\em Phys. Rev. D}
  {\bfseries 94} no.~9, (2016) 094004},
  \href{http://arxiv.org/abs/1509.08226}{{\ttfamily arXiv:1509.08226
  [hep-lat]}}.

\end{thebibliography}\endgroup

\end{document}